\documentclass[a4paper,12pt]{revtex4}

\usepackage{latexsym}
\usepackage{graphics}
\usepackage{color}
\usepackage[latin1]{inputenc}
\usepackage{amssymb}
\usepackage{amsmath}
\usepackage{amsfonts}
\usepackage[dvips]{graphicx}

\usepackage{pdfpages}
\usepackage{epstopdf}

\begin{document}

\title{Low noise phase-locked laser system for atom interferometry}

\author{Bo-Nan Jiang}

\email{bnjiang@siom.ac.cn}

\affiliation{Hefei National Laboratory for Physical Sciences at the Microscale and Department of Modern Physics, University of Science and Technology of China, Hefei, Anhui 230026, China}
\affiliation{Shanghai Branch, CAS Center for Excellence in Quantum Information and Quantum Physics, University of Science and Technology of China, Shanghai 201315, China}
\affiliation{Shanghai Research Center for Quantum Sciences, Shanghai 201315, China}
\affiliation{Present address: School of Science, Shenyang University of Technology, Shenyang, Liaoning 110870, China}

\begin{abstract} A low noise laser system for atom interferometry is realized with phase-locked fiber lasers, where the performance of the OPLL is greatly enhanced by the FEOM feedback loop and the narrow linewidths. The laser system demonstrated contribute 2.2 mrad per shot to the interferometer noise and permit continuous long-term operation for more than 115 hours without relocking in the field test. Also, the mobile gravimeter equipped with this phase-locked laser system reaches a sensitivity as good as 29 $\mu \rm{Gal}/\sqrt{\rm{Hz}}$ and a resolution of 1.1 $\mu \rm{Gal}$ within 1500 s, demonstrating performances comparable to the state of the art.
\end{abstract}

\maketitle

\section{Introduction}

Over the past several decades, atom interferometry has matured to a versatile tool that offers precise and accurate measurement for fundamental and applied sciences\cite{Kasevich,Kai}, especially to inertial sensing\cite{Geiger} such as gravimetry\cite{Peters,Hu,Gillot,Ferier}, gravity gradiometry\cite{McGuirk,Caldani,Bertoldi}, and rotation sensing\cite{Durfee,Dutta}. Instruments of this kind now greatly complement the classical devices\cite{Ferier,Menoret,Wu,Bidel1,Bidel2}. According to their principle of operation, the control of the absolute frequencies and the phase difference of the Raman pulses strongly impact the sensitivity and the accuracy of the measurement\cite{Peters,Gouet,Sorrentino,Gauguet}. Therefore, the laser system for reliable atom interferometric instruments is particularly demanding since it requires both robustness to permit continuous long-term operation and low phase noise, which is critical for improving the measurement sensitivity.

One of the methods to generate Raman pulses is laser modulation with electro-optic modulators\cite{Menoret2,Theron,Luo} or directly applied to the injection current of the laser\cite{Melentiev,Myatt}. In this method, the parasitic sidebands may cause systematic errors in inertial sensing\cite{Carraz} and should be properly handled by tools such as Fabry-Perot cavities\cite{Park} or injection-locking to external grating cavities of diode lasers.\cite{Lu}. In many of the high-accuracy inertial measurements, an alternative method "optical phase-locking" is used to reach the high performance of atom interferometry\cite{Hu,Gillot,Ferier,Menoret,Karcher}. Optical phase-locking loop (OPLL) locks the phases of two independent lasers with an electronic servo loop and has the advantage of being free from the parasitic sideband problem\cite{Santarelli,Cacciapuoti}.

In general, the performance of the phase-locked laser system is limited by the linewidths of the lasers and the bandwidth of the OPLL\cite{Ohtsu}. Thus, in this work, the linewidths of the lasers are further improved to kHz and sub-kHz by applying the fiber lasers, and a novel fiber electro-optic modulator (FEOM) feedback loop is applied to widen the bandwidth of the OPLL. The low noise laser system achieved contribute 2.2 mrad per shot to the interferometer noise and permit continuous long-term operation for more than 115 hours without relocking in the field test. Also, the gravimeter equipped with this phase-locked laser system reaches a sensitivity of 29 $\mu \rm{Gal}/\sqrt{\rm{Hz}}$ and a resolution of 1.1 $\mu \rm{Gal}$ within 1500 s.

\section{Laser setup}

\begin{figure}[h]
\centering
\includegraphics[width=15 cm]{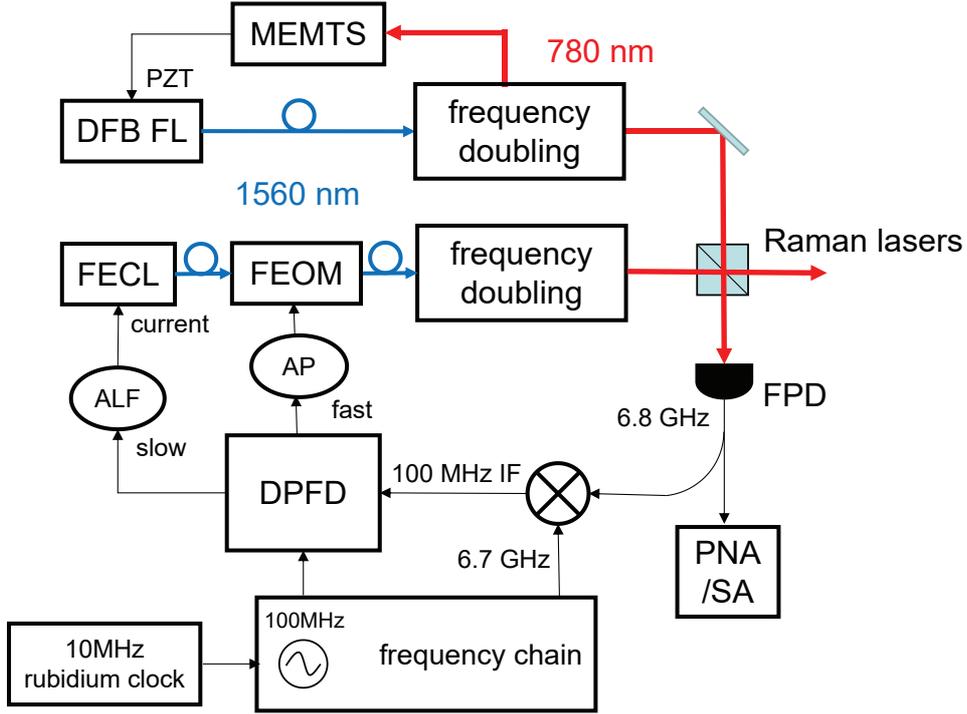}
\caption{Schematic diagram of the experimental setup for the phase-locked laser system. A fast correction is applied on the FEOM, and a slow loop acts on the laser current of the FECL. DFB FL: distributed feedback fiber laser, FECL: fiber external cavity laser, PZT: piezoelectric actuator, MEMTS: magnetic-enhanced modulation transfer spectroscopy, FEOM: fiber electro-optic modulator, FPD: fast photodiode, DPFD: digital phase frequency discriminator, ALF: analog loop filter, AP: analog proportional gain adjust, PNA: phase noise analyzer, SA: spectrum analyzer.}
\label{setup}
\end{figure}

As sketched in Fig. \ref{setup}, our 780 nm laser system for atom interferoemtry is based on the frequency doubling of two 1560 nm seed lasers, one of which is a narrow linewidth (0.83 kHz) distributed feedback fiber laser (DFB FL) with a piezoelectric actuator (PZT) controlled external cavity (NKT Photonics), and the other of which is a narrow linewidth (4.5 kHz) fiber external cavity laser (FECL) with a fixed-length external cavity and a DC-coupled current modulation access (Precilasers). With the magnetic-enhanced modulation transfer spectroscopy (MEMTS)\cite{Long}, the frequency of the master fiber laser is stablized to the $^{87}\rm{Rb}$ D2 transition $|F=1\rangle\rightarrow|F'=0\rangle$, with respect to which the slave fiber laser is red-detuned by 6.8 GHz.

A small fraction (about 250 $\mu$W) of each fiber laser is collected on a fast photodiode (FPD), and the detected beat-note of 6.8 GHz is downconverted by mixing with a 6.7 GHz signal. The 6.7 GHz mixing-down signal is generated by a frequency chain that uses a 100 MHz low phase noise crystal oscillator (LNXO) as a frequency reference. The LNXO is further locked to a 10 MHz signal provided by a rubidium clock to achieve even lower phase noise at low frequencies and to compensate for the long-term drifts. The downconverted beat-note is then compared to the 100 MHz frequency reference from the LNXO with a digital phase frequency discriminator (DPFD, MCK12140), whose output is divided into two paths that provides the error signal for the slow and the fast loops of the OPLL.

For the slow loop, the error signal is sent to an analog loop filter (ALF) which applies a DC-coupled correction on the laser current of the FECL. To reduce the phase noise, the ALF increases the loop gain by using one intermediate-frequency integrator and one low-frequency integrator in series, at the cost of sacrificing the bandwidth. To compensate the reduction of the bandwidth, a FEOM fast loop is added, which feeds the AC-coupled error signal directly back to the FEOM placed in the path of the slave fiber laser. Compared to the correction applied on the laser current or the external cavity, the FEOM feedback loop reaches much larger loop gain and disturbs no laser frequency. An analog proportional gain adjust (AP) tunes the fast loop gain against the ALF loop gain to trade off the residual phase noise and the bandwidth, optimizing the performance of the OPLL.

\section{Residual noise of the optical phase-locking loop}

\begin{figure}[h]
\centering
\includegraphics[width=15 cm]{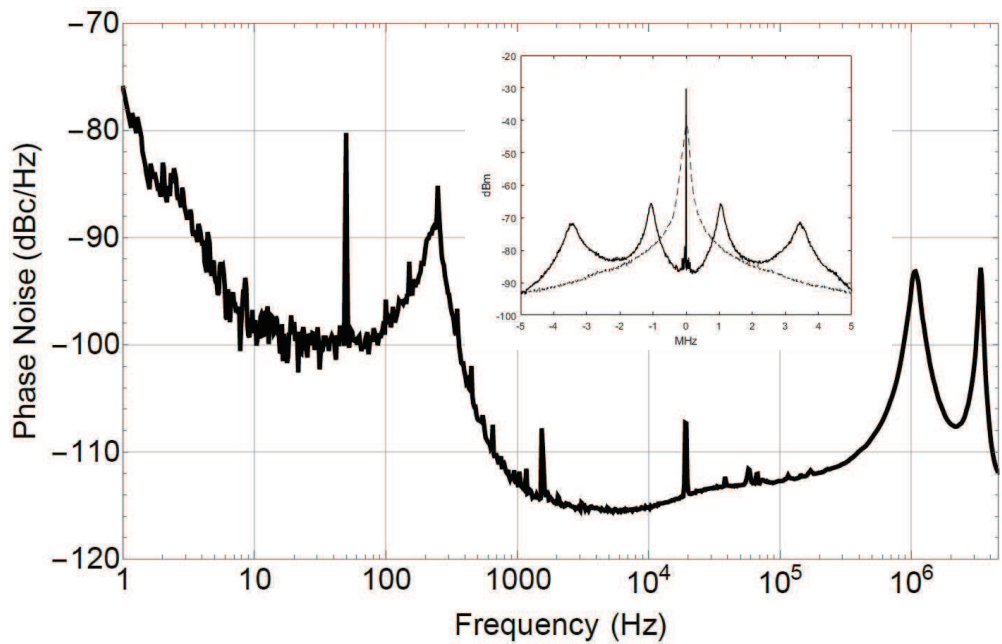}
\caption{Measured phase noise power spectral density of the beat-note signal, with the OPLL being optimized to reduce the phase noise. Inset: The corresponding power spectrum, measured on a SA (Keysight N9020A) with 1 kHz resolution bandwidth, for the optimized OPLL (black solid) and for the free running lasers (gray dashed).}
\label{phasenoise}
\end{figure}

In order to quantitatively characterize the performance of the OPLL, in Fig. \ref{phasenoise}, we measure the power spectral density of the phase-locked beat signal with PNA (Keysight E5052B). The phase noise at low frequency (below 100 Hz) is dominated by the LNXO, except a line-noise-induced excitation around 50 Hz. Between 100 Hz and 300 Hz, the phase noise bumps up due to the vibration noise picked up by the moving components of laser cavities. Higher than 300Hz, the phase noise dives under -100 dBc/Hz and reaches a phase noise level close to -115 dBc/Hz from 1 kHz to 100 kHz. The first servo bump up is observed near 1 MHz, and it is mainly due to the bandwidth of the slow loop, which has a high gain; the second servo bump appears between 3 MHz and 4 MHz, and it is mainly due to the bandwidth of the fast loop, which is limited by the optical delays between the FEOM and the FPD. The phase noise beyond the second servo bump is not corrected by the OPLL anymore and corresponds to the free running laser.

The contribution of the laser phase noise to the interferometer noise can be calculated as\cite{Cheinet}

\begin{eqnarray}
\sigma^{\rm{rms}}_{\phi} = \sqrt{\int^{\infty}_{0}|H_{\phi}(f)|^2 S_{\phi}(f)df},
\end{eqnarray}
with $S_{\phi}(f)$ being phase noise power spectral density of the phase-locked lasers and $H_{\phi}(f)$ being the transfer function; and the corresponding sensitivity limit can be evaluated as\cite{Gustavson}

\begin{eqnarray}
\frac{\sigma^{\rm{rms}}_{\phi}}{k_{\rm{eff}}(T+2\tau)(T+\frac{4\tau}{\pi})\sqrt{\frac{1}{T_{\rm{rep}}}}},
\end{eqnarray}
where $k_{\rm{eff}}$ is an effective wavenumber, $T$ is a time interval between two consecutive Raman pulses, $\tau=$ is the $\pi/2$ Raman pulse width, and $\frac{1}{T_{\rm{rep}}}$ is a repetition rate of measurement cycles. For our mobile gravimeter with $T=$82 ms, $\tau=$12.4 $\mu$s and $T_{\rm{rep}}=$0.34 s, this phase-locked laser system contributes 2.2 mrad per shot to the interferometer noise and the corresponding sensitivity limit is 1.1 $\mu \rm{Gal}/\sqrt{\rm{Hz}}$. And for the transfer function of the ultrahigh-sensitivity gravimeter in Ref.\cite{Hu}, this phase-locked laser system contributes 1.9 mrad per shot to the interferometer noise and the corresponding sensitivity limit is 0.013 $\mu \rm{Gal}/\sqrt{\rm{Hz}}$.

\section{Field experimental results}

\begin{figure}[h]
\centering
\includegraphics[width=15 cm]{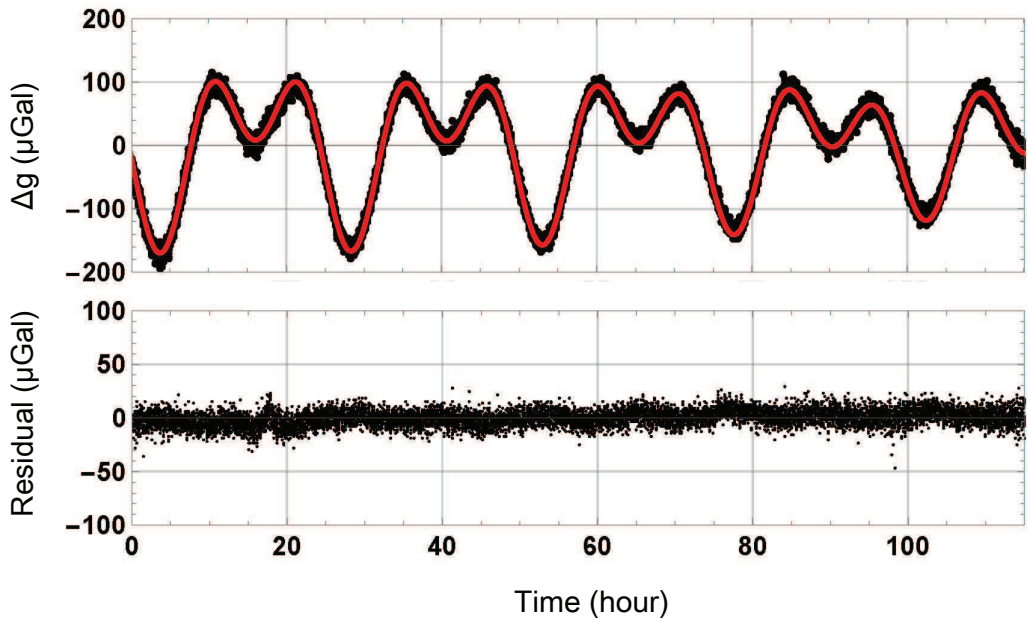}
\caption{Top: the gravity acceleration $\Delta$g measured by a graviemter using this phase-locked laser system between the 30th December 2020 and the 4th January 2021. The setup works continuously for more than 115 hours without relocking the OPLL. Bottom: the residue achieved from the corresponding gravity signal subtracted by Earth$^{\prime}$s tides.}
\label{tide}
\end{figure}

\begin{figure}[h]
\centering
\includegraphics[width=15 cm]{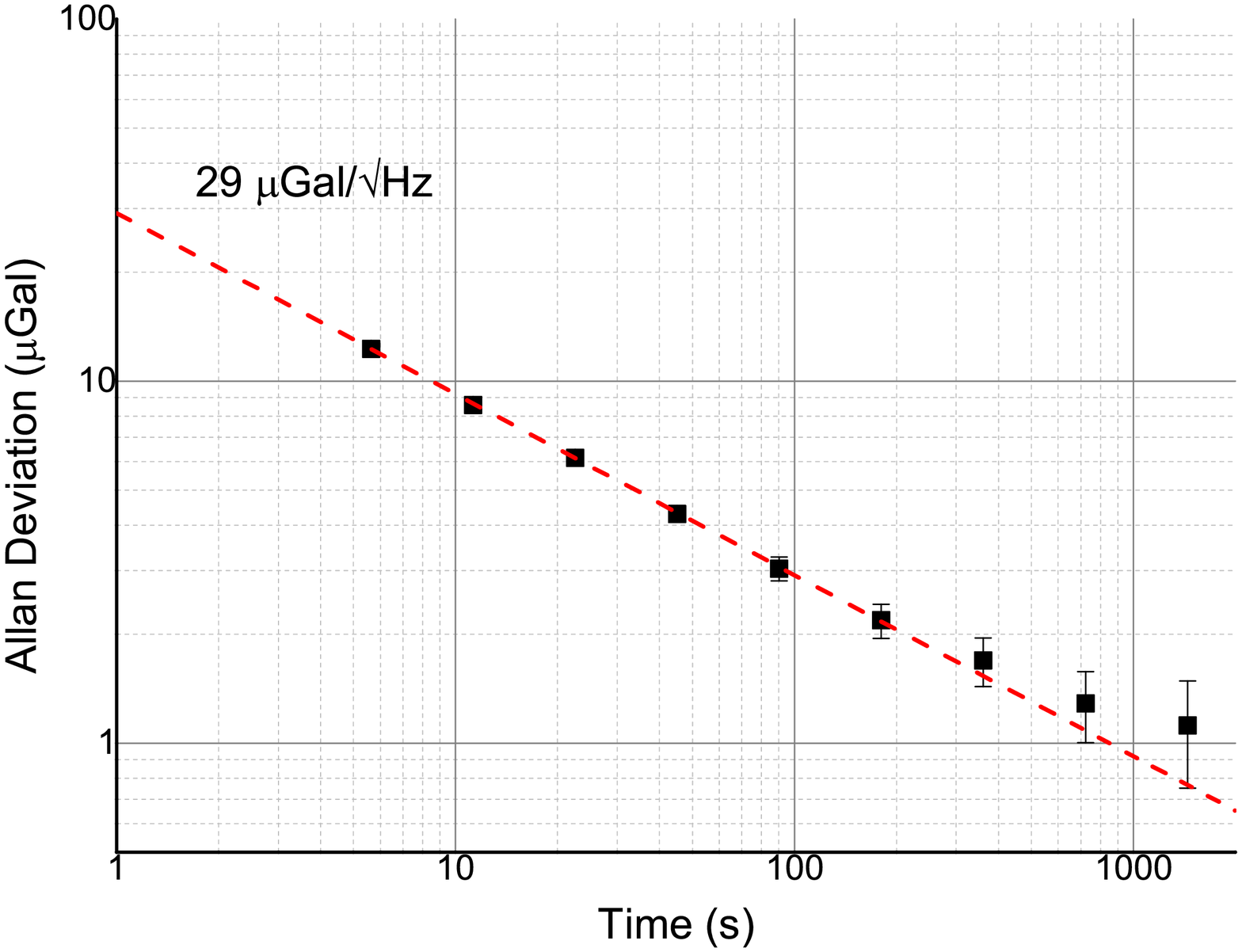}
\caption{Allan deviation of the gravity signal corrected for Earth$^{\prime}$s tides. The $\tau^{-1/2}$ slopes (red) represent the corresponding averaging expected for white noise.}
\label{allan}
\end{figure}

The phase-locked laser system is then integrated into the controller which provides lasers for the portable atomic gravimeter and performs data acquisition and processing. And the mechanically noise of its cooling fans is isolated from the optics by sound insulation. The portable atomic gravimeter is similar to the one described in Ref. \cite{bin}, except that the seismometer correction is applied instead of the active vibration isolation. The test mass is a free falling cloud of Rubidium 87 atoms. The cold atoms are trapped and cooled by a three dimensional magneto-optical trap to a temperature of 3.7 $\mu$k.  After a stage of state preparation and velocity selection, about 10$^{6}$ atoms in magnetic field insensitive state $|F=1,m_{F}=0\rangle$ are obtained. In the process of interference, the free fall atomic cloud is interrogated by a $\pi$/2-$\pi$-$\pi$/2 Raman pulse sequences, the Doppler frequency shift is compensated by chirping the frequency of the Raman beams by a rate of $ \alpha\sim$ 25.1 MHz/s. We obtain the interferometry fringe by changing the chirping rate $\alpha$ with a constant step and finally readout the $g$ value by mid-fringe fitting.

After the completion of the assembly, the gravimeter equipped with the phase-locked laser system traveled 834 km to a seismic station where the field experiment was carried out. The vibration noise of the test site was estimated to be 124$\mu \rm{Gal}$/shot, mainly due to the construction activity underground. We located the gravimeter and the seismometer at the same gravity pillar, thus the vibration noise can be corrected through the seismometer correction\cite{legouet}.

As shown in Fig. \ref{tide}, the gravimeter performs there the continuous measurement of the local gravity for more than 115 hours without relocking the OPLL. And the experimental results (black) agree well with Earth$^{\prime}$s tides (red) predicted theoretically with an inelastic non-hydrostatic Earth model\cite{Dehant}. The Allan deviation of the residue signal is then calculated to characterize the sensitivity and the stability of the gravimeter. As shown in Fig. \ref{allan}, the sensitivity of the gravimeter follows 29 $\mu \rm{Gal}/\sqrt{\rm{Hz}}$ for up to 1000 s. The 10 $\mu \rm{Gal}$ level is obtained within 10 s of measurement, and the Allan deviation continues to decrease down to 1.1 $\mu \rm{Gal}$ with an integration time of 1500 s.

\section{Conclusion}
In conclusion, a low noise laser system realized with phase-locked fiber lasers is demonstrated, where the performance of the OPLL is greatly enhanced by the FEOM feedback loop and the narrow linewidths. The laser system demonstrated contribute only 2.2 mrad per shot to the interferometer noise and are robust enough to permit continuous long-term operation after a long-range transportation. The mobile gravimeter equipped these phase-locked laser system reaches in the field test a sensitivity as good as 29 $\mu \rm{Gal}/\sqrt{\rm{Hz}}$ and a resolution of 1.1 $\mu \rm{Gal}$ within 1500 s. The technique demonstrated here would help us to push the atom interferometry to field applications, such as mobile gravity survey or inertial navigation\cite{Kai,Geiger}.

{\bf Acknowledgments}

The author thanks Prof. Shuai Chen and his team for the discussions and the work in the field test, and also thanks Precilasers team for their technical support. This work is funded by the Youth Program of National Natural Science Foundation of China (Grant No. 11804019), and also supported by the National Key R\&D Program of China (Grant No. 2016YFA0301601), National Natural Science Foundation of China (Grant No. 11674301), Anhui Initiative in Quantum Information Technologies (Grant No. AHY120000), and Shanghai Municipal Science and Technology Major Project (Grant No. 2019SHZDZX01).

Recently, the author became aware that Prof. Franck Pereira Dos Santos and co-workers were also investigating the use of narrow-linewidth fiber lasers and FEOM in a broadband OPLL for atom interferometry\cite{Santos}.

\bibliographystyle{unsrt}
\bibliography{bibliofp}

\end{document}